\documentstyle[prb,aps,preprint]{revtex}
\begin{document}
\draft
\title{In-plane optical response of Bi$_2$Sr$_2$CuO$_6$}
\author{A. A. Tsvetkov$^{a,b}$, J. Sch\"utzmann$^{a}$, 
J. I. Gorina$^{b}$, G. A. Kaljushnaia$^{b}$
and D. van der Marel$^{a}$}
\address{
$^{a}$Material Science Center, Laboratory of Solid State Physics, 
University of Groningen,
Nijenborgh~4, 9747 AG Groningen, The~Netherlands,\\ 
$^{b}$P. N. Lebedev Physical Institute, Russian Academy of Sciences,
Leninsky prospect 53, 117924~Moscow, Russia }
\date{\today}
\maketitle 
\begin{abstract} 
We report on infrared reflectivity measurements of the $ab$-plane response 
of superconducting Bi$_2$Sr$_2$CuO$_6$ single crystals. 
The frequency dependent conductivity has a maximum near 500 cm$^{-1}$ at
room temperature, which shifts to lower frequency and merges with a
Drude peak below 100 K. We attribute the unusual behavior of the
mid-infrared conductivity to low frequency transitions between 
electronic bands of mainly BiO character near the $\overline{M}$ point. 
The linear temperature dependence of the low-frequency 
resistivity can be followed down to approximately 40 K where it 
saturates.
\end{abstract}

%
\narrowtext


The relatively simple crystal structure of single layer compounds offers 
the possibility to study intrinsic properties of the CuO$_2$-plane. 
It is however not clear to what extent the BiO-layers contribute 
to the optical conductivity, and how the hole-doping into the 
CuO$_2$-planes proceeds. 
Previously, it was suggested from an optical study of Tl$_2$Ba$_2$CaCu$_2$O$_8$ 
that the TlO-layers exhibit non-conducting behavior.\cite{jehl} 
The situation for the Bi$_2$Sr$_2$CuO$_6$-system might be more complex. 
Band structure calculations using the generalized potential 
augmented plane-wave (LAPW) 
method by Singh and Pickett \cite{singhpickett} show that 
weak structural distortions 
shift the Bi-O(3) derived bands below the Fermi level. 

Since the normal state dynamics is of special interest for understanding 
strongly correlated electron systems, the low T$_c$ ($< 10$ K) of 
the single layer Bi$_2$Sr$_2$CuO$_6$ compound 
offers the opportunity to examine these 
features well below the transition temperature of the related 
double and triple layer systems. 
It is widely accepted that the in-plane response of the cuprates is 
in the clean-limit and the strong mid-infrared response is due to
strong correlation effects in the cuprates.
In particular the marginal Fermi-liquid (MFL)\cite{varma}, 
and Luttinger liquid\cite{luttinger} approaches along with the
conventional strong electron-phonon interaction\cite{dolgov}
have been put forward to account for these unusual properties.
In these models the strong mid-infrared conductivity 
is related to the linear frequency 
dependence of the scattering rate 
of the quasi-particles in a single band picture. 
Therefore, the normal state properties of the in-plane conductivity 
of Bi$_2$Sr$_2$CuO$_6$ down to approximately 10 K should 
play a pivotal role in a clear separation of the different 
excitations.

In this paper we report on the far-infrared $ab$-plane conductivity of 
a Bi$_2$Sr$_2$CuO$_6$ single crystal that was calculated via 
a Kramers-Kronig analysis of the measured reflectivity data. 
Above 70 K the conductivity increases with frequency until
it reaches a maximum, the frequency of which 
increases with increasing temperature. 
The temperature dependence of the low-frequency conductivity is consistent 
with the $dc$ resistivity measured on crystals of the same batch. 
Based on the f-sum rule we argue that phonon-contributions
are too weak to account for this maximum. We
discuss a scenario, where the unusual temperature dependence is due to
intra- and interband transitions within and between electronic bands 
near E$_F$ of CuO$_2$ and BiO character.


Bi$_2$Sr$_2$Cu0$_6$ single crystals were grown from a precursor
dissolved in liquid KCl.\cite{gorina94}
Due to the formation of an enclosed cavity of a few cm$^{3}$ within the melt
samples were obtained as free standing crystalline platelets of sizes up to
1-2$\times$1-2 mm$^2$ and of thickness 5-20~$\mu$m.\cite{martov95}
A constant cavity temperature of 840--850$^\circ$C and a constant 
temperature gradient (2-3~K/cm) were kept within the melt in 
order to provide permanent transport of the precursor to the growth zone, 
which is important for a free growth of homogeneous crystals with a 
flat mirror surface. Such samples require no mechanical or chemical 
polishing prior to the infrared reflectivity measurements.
X-ray measurements revealed a perfect crystal structure 
with lattice parameters $a=5.36$ \AA\, $b=$5.37 \AA\  and $c=24.64$ \AA, 
which correspond to the orthorhombic $\sqrt{2} \times \sqrt{2}$ 
distortion of the body centered tetragonal structure.
No traces of other phases could be detected.
The rocking curves have a full width at half maximum 
of 0.1$^\circ$,\cite{martov95} which is the minimum value so far reported.
Both from the resistivity and the $ac$-susceptibility we obtained
the same values for the superconducting transition 
temperature (T$_c$=7--8 K) and for the transition width ($\Delta$T$_c$=2 K). 
Unlike earlier reports where a linear resistivity was observed down to
the superconducting phase transition\cite{martin,hou}, in the present crystals
$\rho(T)$ is only linear at high temperatures 
but saturates below 40 K\cite{ved95} 
at a residual resistivity with a sample to sample variation 
from 150 to 300 $\mu\Omega$cm.


The reflectance measurements were made on the a-b plane of 
Bi$_2$Sr$_2$Cu0$_6$ single crystals at normal incidence
for temperatures from 300 K down to 10 K.
We used two Fourier-transform spectrometers to cover the frequency range 
from 50 cm$^{-1}$ to 12000 cm$^{-1}$ and a flow cryostat 
for the temperature variation.
Absolute reflectivities were obtained by referencing the reflected 
intensity of the sample to a Au-mirror with the same shape and size at
each temperature. The systematic errors introduced due to the 
sample/reference interchange fell within 0.5\%.
A further confirmation of the absolute accuracy reached with this
procedure comes from the small ($\approx$ 1 \%) 
mismatch in the absolute reflectivity in 
the overlap region between the FIR and MIR regions.  
 

The temperature dependence of the reflectivity is shown in Fig.~\ref{refl}. 
Decreasing the temperature from 300 K down to 100 K the reflectivity shows 
an increase up to mid-infrared frequencies. 
From 100K downto 40 K the only increase in reflectivity 
occurs in the far-infrared range. The reflectivity is 
temperature independent below 40 K.
The near-infrared reflectivity has a broad plasma minimum 
around 8700 cm$^{-1}$.
Weak reproducible structure is observed at 181, 327, 427, and 508 cm$^{-1}$.
The minima at 327 and 427 cm$^{-1}$ coincide with 
the frequencies of the $c$-axis LO phonon modes of 
Bi$_2$Sr$_2$CuO$_6$\cite{marel}. Under ideal conditions (large perfect
crystals with flat surfaces, perfectly s-polarized plane waves) leakage of
c-axis longitudinal optical phonons into the ab-plane response can be 
excluded\cite{comment}, but some weak and surface dependent
leakage may occur if the experimental conditions are less than perfect.  
As the minima at 181 and 508 cm$^{-1}$ are not seen for $\vec{E}\parallel c$,
we attribute these to in-plane optical phonons.

To calculate the infrared conductivity through a Kramers-Kronig 
transformation we used a Hagen-Rubens extrapolation for the low-frequency 
region. Extrapolation towards high frequencies (up to 320,000 cm$^{-1}$) was 
done using the data of Terasaki {\em et al.}. \cite{tajima90} 

The real part of the conductivity $\sigma_1$ is shown up to 8000 cm$^{-1}$ 
in Fig.~\ref{sigma} and in more detail up to 2000 cm$^{-1}$ in 
Fig.~\ref{2cmpnd}. 
The conductivity at low temperatures is almost featureless 
and decreases monotonically with frequency. 
When the temperature is increased to 100 K, a maximum in the 
conductivity appears, which shifts to 500 cm$^{-1}$ at room temperature. We
also notice from the way the conductivity curves cross, 
that at least part of the spectral weight 
removed from the low frequency side of the spectrum is recovered
in the mid infrared range. To demonstrate this, we present 
in Fig. 4 the carrier density obtained from integrating $\sigma(\omega)$
up to a cut-off frequency 
($n_{eff}(\omega_c)=2m\pi^{-1}e^{-2}\int_0^{\omega_c}\sigma(\omega')d\omega'$). 
The $dc$ resistivity obtained by extrapolation of the optical conductivity to
zero frequency is presented in the insert of Fig. 2. This shows 
a linear temperature dependence, which saturates below 40 K.

The fact that the single layer compounds have almost the same volume 
per unit cell as the double and triple layer compounds, results in 
a low free carrier concentration and a relatively low conductivity. 
Thus, one might expect that the contribution of low-frequency 
excitations, in particular optical phonons, can no longer be neglected. 
However, by applying the $f$-sum-rule  
($\int_0^\infty 8\sigma_{ion}(\omega) d\omega = 
\sum_j 4 \pi n_j Z_j^2/ m_j$) to all ions with mass $m_j$
and ionic charges $Z_j$, we calculate an average
value of the conductivity due to phonons of only 
18 $\Omega^{-1}$cm$^{-1}$ in the range of 0 to 
700 cm$^{-1}$, {\em i.e.} one order of magnitude smaller than
the rise in conductivity between 0 and 500 cm$^{-1}$ in Fig. 2. 
Hence the phonon contributions are far too weak to be
the main source of this rise in conductivity. 

The maximum near 500 cm$^{-1}$ (indicated with arrows in Fig. 3) 
shifts to a lower frequency upon reducing the temperature, and is superimposed 
on a steeply falling free-electron conductivity which increases as the 
temperature is decreased. At 10 K the free-electron response is prevalent 
in the low frequency conductivity. In the remainder of this paper we will
indicate the depression of the conductivity below the maximum as a pseudo-gap.
In the present case there is no unambiguous -and physically meaningful- way 
to decompose the conductivity in free-carrier and bound-charge
components. One of the reasons is, that in the high T$_c$
cuprates also the intraband conductivity deviates considerably from
standard Drude behavior. In particular the high T$_c$ cuprates, if optimally 
doped, are known to have a linear frequency dependency of the scattering rate, 
as follows from inversion of the real and imaginary part of $\sigma(\omega)$ 
using the expression\cite{jimallen}
$\sigma(\omega) = 
 n e^2 m^{-1}/\left\{\gamma(\omega)-i\omega m^{\star}(\omega)/m\right\}$,
where $n$ is the carrier density, $m$ is the band mass, 
$m^{\star}(\omega)/m$ is the effective mass enhancement, and $1/\gamma(\omega)$
is the effective relaxation time. A further complication arises,
if two or more electronic bands cross the Fermi level. In this case, which we 
believe to be relevant for Bi$_2$Sr$_2$CuO$_6$, the spectrum of 
$\sigma(\omega)$ contains interband transitions in addition to the intraband
components. The interband part has a complex line-shape which depends on the 
details of the band-dispersion and the k-dependent optical transition matrix 
elements. Obviously if the material has
two or more bands of electrons, at least one of which (the CuO$_2$ bands) has a 
frequency dependent scattering rate, it is no longer possible to make a 
meaningful separation in several components. At low temperatures, where the 
CuO$_2$ channel has a reduced scattering rate, $\sigma(\omega)$ is dominated by
the CuO$_2$ channel at least at the low frequencies. In this region one
might hope to extract the frequency dependent scattering rate of the CuO$_2$
carriers, at least up to a frequency where other contributions to 
$\sigma(\omega)$ become significant.
For this reason we display in the inset of Fig. 4 the frequency dependent
scattering rate only for the lowest temperature. We notice that 
the linear behavior of $\gamma(\omega)$, which is rather typical 
for all high T$_c$ cuprates, exists only up to 0.1 eV in this case. 
In view of the possibility of low energy interband transitions in this compound, 
even at 10 K $\gamma(\omega)$ extracted from optical data is probably no 
longer meaningful for $\hbar\omega > 0.1$ eV. 
Our observations support the conclusion by Romero $et$ $al$, 
based on an analysis of infrared transmission data, that the linear 
frequency dependency of $\gamma(\omega)$ is limited to the energy range 
below 0.1 eV.\cite{romero}
The non-linear temperature dependence of the resistivity in
the crystals used for this study is also inconsistent 
with MFL behavior at the lowest temperatures. 

Transmission experiments on free standing non-superconducting single crystals 
of Bi$_2$Sr$_2$CuO$_6$ performed by Romero {\it et al}\/\cite{romero} 
also reveal a singularity in the FIR conductivity, although the effect is less 
pronounced. The room temperature conductivity shows a clear change of slope 
around 500 cm$^{-1}$, and the frequency dependent effective mass 
$m^\star(\omega)$ has a minimum at this frequency. 
While only a trace of the pseudo-gap is observed in 
non-superconducting single crystals\cite{romero}, the effective number
of carriers remains the same. 
A maximum in the optical conductivity was also observed in the related single 
layer compound Tl$_2$Ba$_2$CuO$_6$,\cite{puchkov} although a substantially 
larger spectral weight was involved.

A possible candidate for the observed pseudo-gap is the additional 
electronic degree of freedom introduced by having both hole-doped CuO$_2$ 
and electron-doped BiO bands. 
Although an interpretation in terms of direct transitions
between these bands is probably too simple due to the strong electron-electron
correlations in the Cu-O bands, a pseudo-gap may nevertheless 
exist, not unlike the situation encountered 
in Kondo-insulators.\cite{kondo-insulators}
The LAPW band calculations of Singh and Pickett\cite{singhpickett} give 
two types of bands: For a vanishing orthorhombic distortion there is
a large Cu-O derived barrel section centered at X and its
$\sqrt{2}\times\sqrt{2}$ folded counterpart centered around $\Gamma$. 
In addition there are two Bi-O derived electron pockets with
their BZ-folded counter parts, all of them centered at
$\overline{M}$. 
As suggested by Singh and Pickett,\cite{singhpickett} a weak structural 
change from tetragonal to orthorhombic 
strongly affects the structure of the electronic 
bands. The bands are shifted up to 400 meV 
for Bi displacements of 0.14 \AA\ and 0.41 \AA\ for O. 
As a result 5 bands were calculated for $k$ close to the  
saddle point ($\overline{M}$), 
all within a range of 200 meV around E$_F$. Only
two of the Bi-O pockets (one of them barely) cross $E_F$ when the 
orthorhombic distortions are taken into account. 
Because the band structure is strongly affected by modulations
of the lattice, it can also have an appreciable temperature dependence.
Photoemission studies have revealed a single occupied band near E$_F$ 
close to the $\overline{M}$-point, which was originally indicated as a possible
BiO derived band\cite{ratner} but was identified as a CuO$_2$ band in a
later publication\cite{king}. The absence of a clear BiO-derived band
in photoemission may however result from a disorder induced
smearing of the $k$-dispersion in the BiO-layers, or from the 
small photoelectron cross section for Bi and O states compared to the
Cu 3d cross-section. The main 'generalized Drude' component in our
data then originates from the CuO barrels, with a linearly
frequency dependent decay-rate typical for optimally 
doped cuprates. The pseudo-gap at high temperatures would be
due to transitions between the closely spaced bands 
which cross E$_F$ near $\overline{M}$. 
In this scenario the disappearance of 
the pseudo-gap at low temperatures can reflect a gradual change
of atomic coordinates upon cooling, with a corresponding change
of the interband separation. This structural modulation can easily
result in a re-distribution of carriers between BiO and CuO$_2$ bands and 
the disappearance of the pseudo-gap. Interestingly, large lattice distortions 
are generally present in films, which show a higher T$_c$ and a more 
pronounced pseudo-gap.\cite{pascaleroy} 


We have presented experimental data on the in-plane infrared response of 
superconducting Bi2201 single crystals for temperatures from 10 K to 300 K. 
The optical conductivity shows a pseudo-gap feature, that shifts with 
temperature. 
The optical conductivity is at least one order of magnitude too large to be 
compatible with a purely phononic interpretation of this feature. 
This observation indicates the presence of two coupled conducting channels
and may provide important clues regarding the fact that T$_c$ in the single 
layer Bi-cuprate material is much lower than in other nearly stoichiometric 
compounds. We discuss the possibility that the pseudo-gap is related to 
electronic transitions between bands near the $\overline{M}$-point, which
are strongly coupled to the orthorhombic lattice distortion.

We gratefully acknowledge many useful comments from W. N. Hardy.
This investigation was supported by the program for Russian-Dutch 
Cooperation under project \# 047003033 of the Nederlandse Organisatie 
voor Wetenschappelijk Onderzoek, the
Russian Scientific Technical Program "High Temperature Superconductivity" 
(A.A.T., J.I.G., G.A.K.), and the program Human Capital and Mobility 
under contract ERBCHBICT941820 (J.S.).


\begin{figure}
\caption{Reflectivity of a Bi$_2$Sr$_2$CuO$_6$ single crystal
at normal incidence and with the electric-field vector parallel to 
the $ab$-plane for T = 10, 100, 200, and 300 K. Inset: Reflectivity 
on an expanded scale for T = 10, 40, 70, 100, 200, 300 K.}
\label{refl}
\end{figure}

\begin{figure}
\caption{Optical conductivity $\sigma(\omega)$ as a function of
frequency for the same temperatures as in Fig.~1.
Inset: $dc$-resistivity obtained by extrapolating the optical
conductivity to zero frequency.
}
\label{sigma}
\end{figure}

\begin{figure}
\caption{Optical conductivity $\sigma(\omega)$ on an expanded scale
for 10, 100, 200, and 300 K. The arrows indicate the evolution of the
maximum.
}
\label{2cmpnd}
\end{figure}

\begin{figure}
\caption{Effective number of carriers per unit of CuO$_2$ calculated from 
$\sigma(\omega)$ using the partial {\it f}--sum rule. 
Inset: Frequency dependent scattering rate ($\gamma(\omega)$) and 
effective mass enhancement ($m^{\star}(\omega)/m$) at 10 K.
}
\label{negam}
\end{figure}

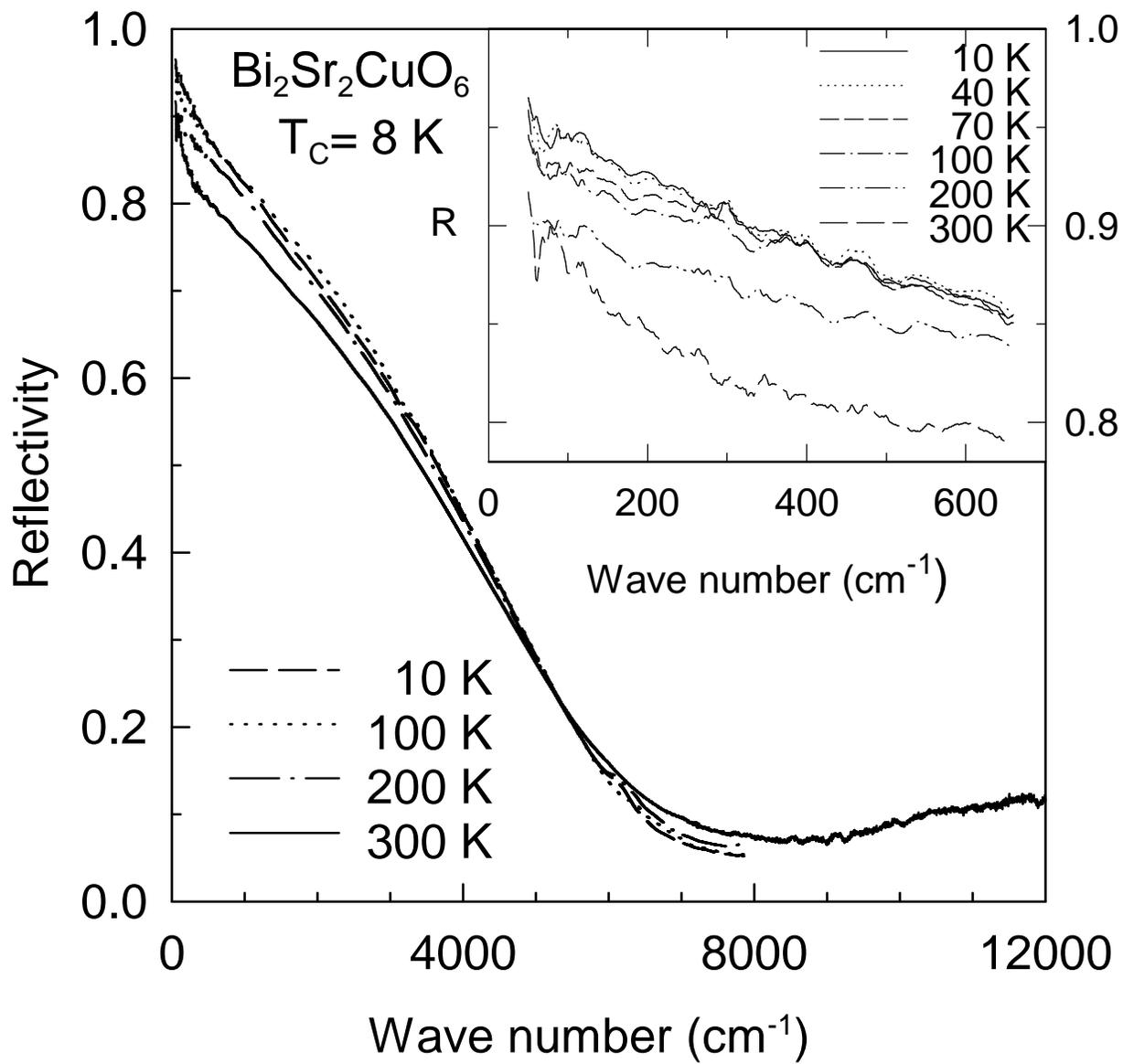
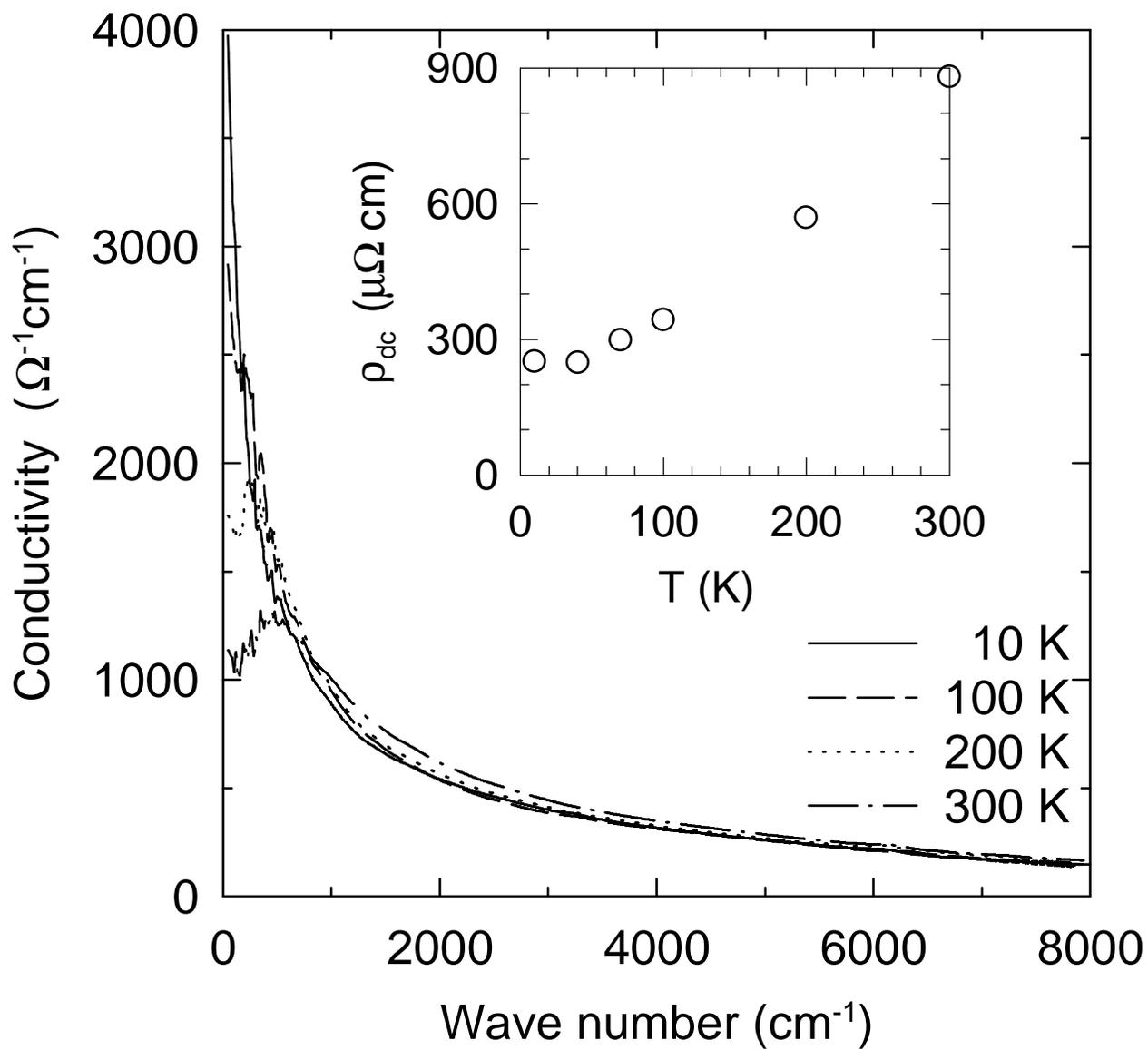
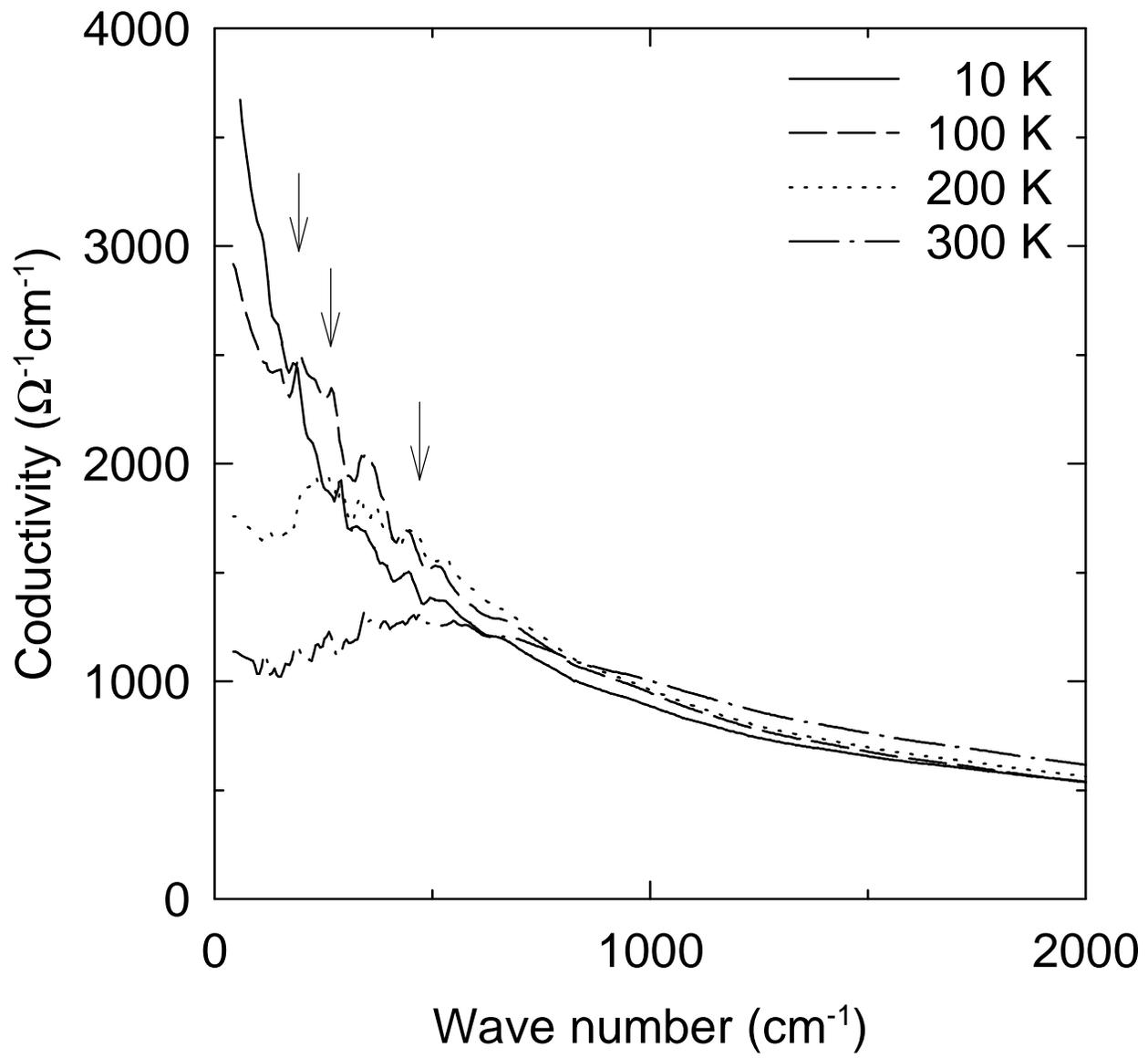
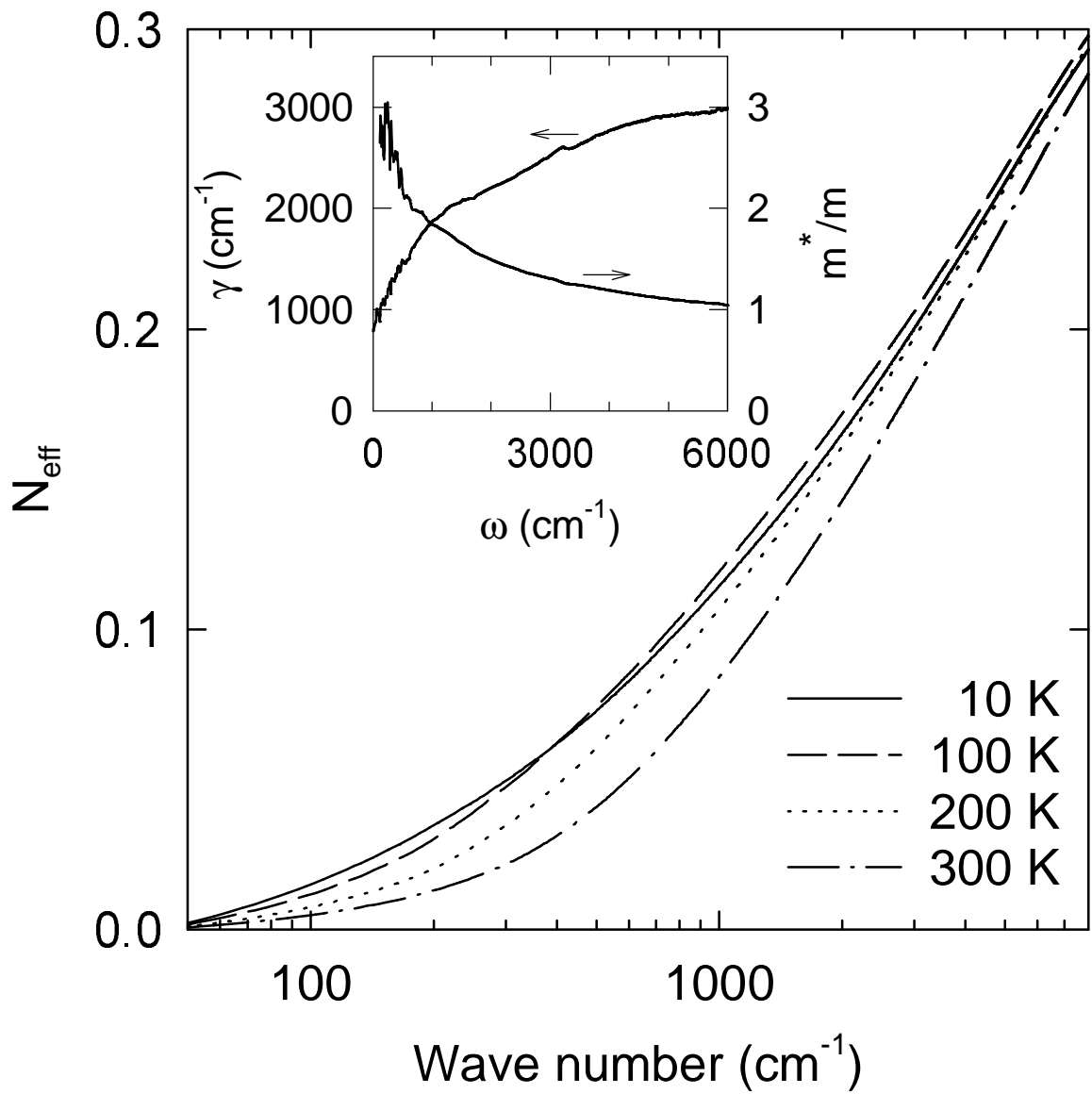
\end{document}